\documentclass[aps,prl,twocolumn,superscriptaddress,showpacs]{revtex4}
\usepackage{epsfig}
\usepackage{graphicx}
\usepackage{bm}%bold math
\begin{document}

\title
{Implementation of quantum operations on single photon {\it qudits}}
\author{Bing He} \email{bhe98@earthlink.net}
\affiliation{Department of Physics and Astronomy, Hunter College of the City University of New York, 695 Park Avenue, New York, NY 10021, USA}
\author{J\'{a}nos A. Bergou  }
\affiliation{Department of Physics and Astronomy, Hunter College of the City University of New York, 695 Park Avenue, New York, NY 10021, USA}
\author{Zhiyong Wang}
\affiliation{School of Physical Electronics, University of Electronic Science and Technology of China, Chengdu 610054, China}

\date{\today}
\begin{abstract}
We show that a general linear transformation from one single photon {\it qudit} to another, the dimension of which can be either equal or unequal to that of the first one, can be implemented by linear 
optics. As an application of the scheme we elaborate a method to deterministically realize any finite-element Positive Operator 
Value Measure (POVM) on single photon signals, which is also generalizable to any quantum system in principle. 
\end{abstract}

\pacs{42.50.Dv, 03.65.Ta, 03.67.-a}

\maketitle
Linear optics is considered as one of the promising candidates for quantum computing (for a recent overview see, e.g., \cite{p-m}) and can be also applied 
to many other areas such as quantum cryptography (for a review see. e.g., \cite{cryto}). In these applications an essential technique is the implementation of all possible operations, including generalized quantum measurements in the form of Positive Operator Value
Measures (POVMs), on the signals encoded as photon states by practical linear optics circuits. 

A typical and important case of the signal states is single photon {\it qudits}, i. e., the linear combinations of the modes $a^{\dagger}_k|0\rangle$, $k=1,\cdots,N$ (multiple-rail encoding). 
It was proposed by Reck {\it et al.} \cite{reck} that any unitary operator ${\cal U}\in U(N)$ on the $N$-dimensional {\it qudits}, $\sum_{i=1}^Nc_ia^{\dagger}_i|0\rangle$, can be realized by an N-port interferometer, which is an array of beam splitters and phase 
shifters performing $SU(2)$ elements, because this unitary operator can be decomposed into the product of these $SU(2)$ elements (see Fig. 1). This scheme was further studied in \cite{torma96,torma95,jex95} and has been
applied to a variety of research fields in quantum information theory and experiment.

\begin{figure}
\includegraphics[width=50truemm]{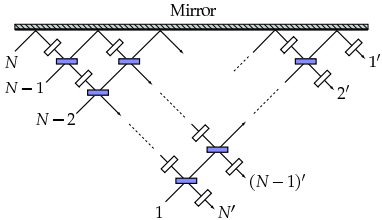}
\caption{The unitary transformation module constructed with beam splitters (dark square) and phase shifters
(white square). Any unitary operator $U$ can be decomposed into the product of $SU(2)$ elements implemented by the beam splitters and the phase shifters, 
and the maximum number of beam splitters needed is $N(N-1)/2$. The input ports are with unprimed numbers and output ports with 
primed numbers.} 
\end{figure}

The generalization of the scheme is the implementation of all possible linear maps on single photon {\it qudits}, which is 
a fundamental task in processing quantum information. It is intimately related to the realization of POVMs that are at the heart of 
many quantum information processing protocols. A finite-element POVM is a set of non-negative operators $\{\Pi_i\}$, where
$\Pi_i$ are its elements, satisfing
\begin{eqnarray}
\sum^n_{i=1}\Pi_i=I,
\end{eqnarray}
with $I$ being the identity operator. 
It has been proved that any rank-one POVM in the form of $\Pi_i=k_i^2|\phi_i\rangle\langle\phi_i|$, where $\langle\phi_i|\phi_j\rangle\neq \delta_{ij}$ and 
$|k_i|\leq1$, can be realized by the Neumark extension \cite{AP93}, which extends the POVM elements to the orthogonal projectors in a larger space. For the input signals prepared with single photons, such a POVM can be implemented with linear optics circuits, performing unitary transformations, and photon detectors only \cite{cal}. 
The realization of POVMs with arbitrary rank is, however, much more difficult. Since $\Pi_i \geq 0$, 
it can be decomposed into $\Pi_i=A^{\dagger}_iA_i$ with $A_i$ being upper triangle matrix \cite{bt}. 
The general POVM will be implemented if we simultaneously realize the maps,
\begin{eqnarray}
\rho_{in}\rightarrow \rho_{out,i}=\frac{A_i\rho_{in} A^{\dagger}_i}{Tr(A_i\rho_{in} A^{\dagger}_i)},
\end{eqnarray}
and successfully detect these outputs. The detection operators $A_i$ (and their transforms by an arbitrary unitary operator $A'_i=U_iA_i$) 
can be any allowed linear map in quantum mechanics with equal dimensional 
input and output signal space, i.e., an $N\times N$ square matrix.

The detection operators of a POVM are, however, only a special case of more general maps called quantum operations (QOs). 
A QO connects pair of input and 
output states via the map,
\begin{eqnarray}
\rho_{in} \rightarrow 
\rho_{out}=\frac{{\cal E}(\rho_{in})}{\mbox{Tr}\bigl({\cal E}(\rho_{in})\bigr)}.
\end{eqnarray}
${\cal E}$ is a linear, trace-decreasing map that preserves the complete positivity (CP), and generally occurs with non-unit probability
$\mbox{Tr}\bigl({\cal E}(\rho_{in})\bigr)\leq 1$. The general form of ${\cal E}$ is given as \cite{bibkraus}
\begin{eqnarray}
{\cal E}(\rho_{in}) = \sum_i K_i \rho_{in} K_i^{\dagger}\;, 
\end{eqnarray}
with the Kraus operators $K_i$ satisfying the bound $\sum_i K_i^\dagger K_i \le I$.
If a QO transforms pure states to pure it is called a pure map. In this case there is only one term 
$K \rho_{in} K^{\dagger}$ in the above equation with $K$ being a {\em contraction}, i.e., $||K||\le 1$.
For a single pure state input $\rho_{in}=|\psi_{in}\rangle\langle\psi_{in}|$ \cite{input}, the QOs in Eq. (3) can be therefore written in the form \cite{d-p}, 
\begin{eqnarray}
|\psi_{in}\rangle\rightarrow \frac{K|\psi_{in}\rangle}{||K|\psi_{in}\rangle||}
\;\label{psimap}.
\end{eqnarray}
The output signal of such a linear map, with $K$ as contraction, is detected with a probability $\langle\psi_{in}|K^{\dagger}K|\psi_{in}\rangle\leq 1$.
These contractions can be more general than the detection operators of a POVM because the input space dimension $N_1$ and the 
output space dimension $N_2$ of $K$ can be different, so $K$ is an $N_2\times N_1$ matrix whose entries are complex numbers. 
The detection operators $A_i$ of a POVM correspond to a special case of contractions when $N_{1} = N_{2} \equiv N$. 

We now address the problem of how to realize any possible linear map $K$ on single photon {\it qudits} with {\it only} three 
unitary operator modules of the kind shown in Fig. 1. We realize $K$ by its unitary dilation, ${\cal U}$, the unitary operator constructed from $K$ 
in a larger space, which we obtain by using the direct sum extension of the system with an ancilla, 
${\cal H}_{S} \oplus {\cal H}_{A}$. In terms of Hilbert space dimensionality, this scheme minimizes the physical resources 
needed to realize a QO \cite {b-d-s}. We embed the state vector $(c_1,c_2,\cdots,c_{N_1})^T$ (T stands for transpose) of the input signal, $|\psi_{in}\rangle=\sum_{i=1}^{N_1}c_ia^{\dagger}_i|0\rangle$, into a larger space and map it by ${\cal U}$ to a vector containing the state vector of the output, 
$|\psi_{out}\rangle=\sum_{i=1}^{N_2}c'_ia'^{\dagger}_i|0\rangle$ (unnormalized), of $K$:
\begin{eqnarray}
\left(\begin{array}{c}c'_1\\
 \vdots\\
 c'_{N_2}\\
  \vdots 
 \end{array}\right)=\left(\begin{array}{cccc}{\cal U}_{1,1}&{\cal U}_{1,2}&{\cal U}_{1,3} &\cdots \\
{\cal U}_{2,1} & {\cal U}_{2,2}&{\cal U}_{2,3} &\cdots \\
{\cal U}_{3,1} &{\cal U}_{3,2}&{\cal U}_{3,3}& \cdots\\
  \vdots&\vdots &\vdots & \ddots
\end{array}\right)\left(\begin{array}{c}c_1\\
 \vdots\\
 c_{N_1}\\
 \vdots\\
 0
\end{array}\right).~~~~
\end{eqnarray}
It should be noted that we realize this unitary dilation always with a vacum state ancilla $(0,0,\cdots,0)^T$ added to the input state vector as a direct 
sum. Next, we will derive the algorithm to generate the unitary dilation ${\cal U}$ of linear map $K$. 

$K^{\dagger}K$ and $KK^{\dagger}$ are positive matrices of $N_1\times N_1$ and $N_2\times N_2$, respectively.
Suppose $N_1 \geq N_2$, we choose to diagonalize $KK^{\dagger}$ by a unitary operator $U$.
We can also construct an $N_1\times N_1$ unitary matrix $V$ to obtain the Singular Value Decomposition (SVD), 
$K=U\Sigma V^{\dagger}$, with the uniquely determined {\it singular values} on the diagonal of the $N_2\times N_1$ matrix $\Sigma$. 
If $N_1< N_2$, on the other hand, we will choose to diagonalize $K^{\dagger}K$ to get a similar result.

Then we extend the rectangular matrix $\Sigma$ to the following
$max(N_1,N_2)\times max(N_1,N_2)$ square matrix
\begin{eqnarray}
\Sigma'=\left(\begin{array}{cccccc}|\sigma_1| & & & & &  \\
 &\ddots & & & &\\
& & |\sigma_{N_2}|& & &\\
 & & & 1 & &\\
 & & & & \ddots &\\
 &  &  &  &  &  1
\end{array}\right),
\end{eqnarray}
where the singular values $\sigma_i$ satisfy $|\sigma_i| \leq 1$ since $K$ is a contraction.
The extension of $\Sigma$ in the case of $N_1< N_2$ also takes the above form except that $|\sigma_{N_2}|$ is replaced
by $|\sigma_{N_1}|$.
Using the fact that $\Sigma'$ is still a contraction, we can obtain a $max(2N_1, 2N_2)\times max(2N_1, 2N_2)$ unitary 
dilation 
\begin{eqnarray}
G=\left(\begin{array}{cc}\Sigma'& (I-\Sigma'^2)^{1/2}\\
 (I-\Sigma'^2)^{1/2}& -\Sigma'\\
\end{array}\right)
\end{eqnarray}
of it (see, e.g., Ex. I.3.6 in \cite{bt}), which acts on a space $\mathcal{H}\oplus \mathcal{H}$
with $\mathcal{H}$ being $max(N_1, N_2)$ dimensional. 
We also extend $U$ and $V$ to $max(2N_1, 2N_2)$ by $max(2N_1, 2N_2)$ matrices by adding the identity matrix $I$ in the diagonal and zero matrices off the 
diagonal.
A general linear map $K$ is therefore realized by the following 
unitary dilation:
\begin{eqnarray}
{\cal U}=UGV^{\dagger}.
\end{eqnarray}

In our setup, we perform its equivalence by acting ${\cal U}^{\dagger}$ on the spatial mode vector $(a^{\dagger}_1,\cdots,a^{\dagger}_{N_1})$ .
The circuits to implement $V$ and $U^{\dagger}$ are the corresponding $N_1$-port and $N_2$-port modules.
After the input spatial mode vector is processed by $V$, we redirect
the output to a $max(2N_1, 2N_2)$-port module of $G$ with the input ports numbered from $N_1+1$ to $max(2N_1, 2N_2)$ in Fig. 1 black 
or a vacum state. Here is some detail about the step to implement $\Sigma$ 
through its unitary dilation $G$. Picking out the entries containing only one of the singular values $\sigma_i$ from the matrix of 
$G$, we form a $2 \times 2$ submatrix, which can be transformed by a rotation 
$T_{i, i+max(N_1,N_2)}$ to a diagonal one:
\begin{eqnarray}
&&\left(\begin{array}{cc}|\sigma_i| & (1-\sigma^2_i)^{1/2}\\
 (1-\sigma^2_i)^{1/2}&-|\sigma_i| \\
\end{array}\right) \left(\begin{array}{cc}\cos\theta_i & -\sin\theta_i\\
 \sin\theta_i &\cos\theta_i\\
\end{array}\right)\nonumber\\
&~~~~~~~ &=\left(\begin{array}{cc}1 & 0\\
0 & -1\\
\end{array}\right).
\end{eqnarray}
With a series of rotations in the form of $T_{i,i+max(N_1,N_2)}\otimes I_{\rm rest}$, for $i=1,\cdots, min(N_1,N_2)$,
$G$ can be realized by $min(N_1,N_2)$ beam splitters with the reflection coefficients $R=1-\sigma^2_i$ and phase shifters giving rise to $e^{i\pi}$. 
Therefore, the upper bound of the total number 
of the beam splitters required in the 
scheme is
\begin{eqnarray}
N_{max}=\frac{N^2_1}{2}+\frac{N^2_2}{2}-|\frac{N_1}{2}-\frac{N_2}{2}|,~~~~~~
\end{eqnarray}
which is determined by the dimensionalities of the input and output Hilbert spaces.  
In the whole extended space, we will obtain two outputs after the action of the three
unitary operator modules: one is the exact output $(a'^{\dagger}_1,\cdots,a'^{\dagger}_{N_2})$ of the linear map $K$ from the output ports of $U^{\dagger}$, 
and the other is an extra output $(a'^{\dagger}_{N_2+1},\cdots,a'^{\dagger}_{max(2N_1, 2N_2)})$ from the output ports of $G$ numbered 
from $(N_2+1)'$ to $\left(max(2N_1, 2N_2)\right)'$. Fig. 2 displays the scheme that realizes the effect of $K$ on the input state $\rho_{in}$.

\begin{figure}
\includegraphics[width=60truemm]{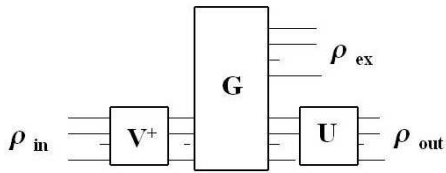}
\caption{The circuit to perform the unitary dilation ${\cal U}$ of $K$, the general linear transformation on a pure state input
$\rho_{in}=|\psi_{in}\rangle\langle\psi_{in}|$. We will obtain two outputs, $\rho_{out}$ of $K$ and an extra
$\rho_{ex}$, from the corresponding terminals. Because the linear map $K$ is not generally inversable, we need to add an 
ancilla $\rho_{ex}$ to $\rho_{out}$,  i.e., $\rho'_{in}=\rho_{out}\oplus\rho_{ex}$, if we are to convert it back to the original $\rho_{in}$ by using the same circuit from the inverse direction.} 
\end{figure}

This linear optics scheme can be directly applied to where we need non-unitary transformation on photon states, e.g., in the production of single photon {\it qudits} 
in any form of $\sum_i c_ia^{\dagger}_i|0\rangle$, where $\sum_i|c_i|^2\leq 1$ 
(possibly unnormalized), by multiple-rail encoding, and in the enhancement of the entanglement of a pair of partially entangled 
photons like $\sum_i c_{i}a^{\dagger}_i|0\rangle_1a^{\dagger}_i|0\rangle_2$, with different $|c_i|$ unequal, by one party operation. 

Now we look in some detail at the realization of POVMs as one important application of our scheme.
We start with the simplest situation of $n=2$, where the two POVM elements are always commutative, $[\Pi_1, \Pi_2]=0$. 
Suppose that the dimension of the signal space is $N$, and the $N\times N$ detection operators $A_i$ of the POVM can be factorized by SVD as $A_i=V_i \Sigma_i U_i$ with $U_i$, $V_i$ unitary and $\Sigma_i$ diagonal. 
We first set up a $N$-port module for $U_1$ and, after the signal leaving $U_1$ module, we process it with a $2N$-port module to implement the unitary dilation of $\Sigma_1$. From its output ports  
numbered from $1'$ to $N'$ we get an output $|\psi^1_{mid}\rangle\sim \Sigma_1U_1|\psi_{in}\rangle$,
while from the ports numbered from $(N+1)'$ to $2N'$ another output $|\psi^2_{mid}\rangle\sim \Sigma_{1C}U_1|\psi_{in}\rangle$, with 
$\Sigma^2_{1C}=I-\Sigma^2_1$. Then we will just redirect them to modules of $V_1$ and $V_2$ and finally obtain the outputs $A_1|\psi_{in}\rangle/||A_1|\psi_{in}\rangle||$ and $A_2|\psi_{in}\rangle/||A_2|\psi_{in}\rangle||$ from the corresponding terminals.

For a POVM with the number of elements $n\geq3$, the situation is much trickier. Instead of $\Pi_2$, what we realize from the 
corresponding output ports of $|\psi^2_{mid}\rangle$ is the operator $I-\Pi_1$. By diagonalization all elements of a general POVM can be factorized into $\Pi_i=U^{\dagger}_i\Sigma^2_iU_i$, where the different $U_i$ do not generally commute, i.e., $[U_i,U_j]\neq0$ for $i\neq j$.
In the realization of $\Pi_2$, therefore, we need to consider two different situations: (1) If $||\Pi_1||<1$ \cite{definition}, because $I-\Pi_1 > \Pi_2$ when $n\geq3$, we can find a diagonal matrix 
$\Sigma^{\ast}_2$ with $||{\Sigma}^{\ast}_2||\leq 1$ \cite{reason1} and a unitary operator $U_{2L}$ such that
\begin{eqnarray}
\Pi_2 &= & U^{\dagger}_1\Sigma_{1C}U^{\dagger}_{2L}\Sigma^{\ast 2}_2U_{2L}\Sigma_{1C}U_1 =
 U^{\dagger}_2\Sigma^{2}_2 U_2.
\end{eqnarray}
These two matrices $\Sigma^{\ast}_2$ and $U_{2L}$ are obtained by a standard digonalization procedure following the above equation. Then, after performing 
$U_{2L}$ with a $N$-port module, $\Sigma^{\ast}_2$ as a contraction map can be implemented by a $2N$-port linear optics module with at most $N$ beam splitters. To realize $A_2$ completely, we add one more module of a proper
$V_2$. (2) If $||\Pi_1||=1$, after the signal goes through the part of circuit implementing $I-\Pi_1$, some of the output ports will be black 
because the corresponding components have been projected out by $\Pi_1$ \cite{reason2}. Then we will just inverse the remaining $(N-D)\times(N-D)$ non-zero part of $\Sigma_{1C}$ matrix, where $D$ is the multiplicity of the unit eigenvalue of $\Pi_1$, in finding 
$U_{2L}$ and $\Sigma^{\ast}_2$ of this size in Eq. (12). 

\begin{figure}
\includegraphics[width=70truemm]{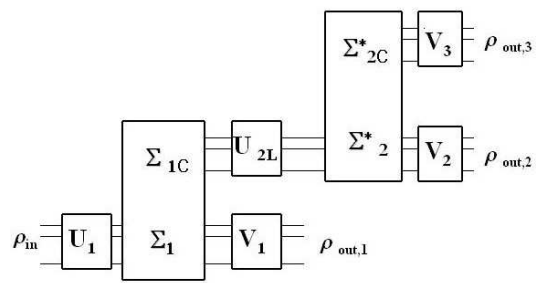}
\caption{The setup to perform any POVM with three elements. The seven unitary operator modules, two of which perform the
pairs of operators $\Sigma_1$ and $\Sigma_{1C}=(I-\Sigma^2_1)^{1/2}$, ${\Sigma}^{\ast}_2$ and $\Sigma^{\ast}_{2C}=(I-\Sigma^{\ast 2}_{2})^{1/2}$, respectively, are designed
with the POVM elements. The detectors at the terminals effect a dephasing to eliminate the interference between 
$A_i|\psi_{in}\rangle$ \cite{L-L}, and capture the outputs $\rho_{out,i}=A_i\rho_{in} A^{\dagger}_i/Tr(A_i\rho_{in} A^{\dagger}_i)$, for $i=1,2,3$, with
the prbabilities $p_i=Tr(A_i\rho_{in} A^{\dagger}_i)$. The outputs redirected to only one set of detectors form a probability distribution as the mixture,
$\sum_ip_i\rho_{out,i}=\sum_iA_i\rho_{in} A^{\dagger}_i$ \cite{watrous}, there.} 
\end{figure}

Repeating the above procedure from the output ports where the operator $I-\Pi_1-\Pi_2$ is realized, we add all the corresponding modules performing $U_{nL}$, $\Sigma^{\ast}_n$, etc., for $n\geq 3$, to implement the remaining 
$A_3$, $A_4$, $\cdots$, $A_n$, respectively. The total number of modules of Fig. 1 needed in our scheme to realize a general POVM with $n$ elements is $3n-2$.
As an illustration of this general scheme, Fig. 3 shows the setup to perform any POVM with three elements.

We have reduced the problem of realizing a POVM to that of finding a sequence of unitary operators with the POVM elements and realizing them with the ancilla states of vacum and 
then detecting the outputs with the standard projective measurements on the extended space. The algorithm to obtain these unitary operators is given, and the implementation of any POVM, with elements of arbitrary rank, 
can be therefore realized for single photon input signals. Using this method, we will directly obtain the output states of a POVM, which can be tailored by choosing the appropriate $V_i$ modules, from the corresponding terminals where the signal detectors are 
placed. If our signals are just single photon polarization qubits, $|\psi_{in}\rangle=c_1|H\rangle+c_2|V\rangle$ (polarization modes $H$ and $V$), we can use much simpler circuit to implement any POVM on them as in \cite {ap}, where a POVM is realized as the decomposition of an identity operator 
but the necessary algorithm to obtain, e.g., $\Sigma^{\ast}_n$, $U_{nL}$ for the implementation of all specified $\Pi_i$ is not given. Given the beyond-linear-optics methods to implement unitary operations on more complicated quantum systems than single 
photon states, this scheme can be applied to more general situations of photonic states as well as the signals of any other type of radiation.

In the whole extended output space ${\cal H}_T$, the overall output of the unitary map ${\cal U}$ implemented by the POVM circuit is 
$|\psi'_{out}\rangle=\left((A_1|\psi_{in}\rangle)^T,\cdots, (A_n|\psi_{in}\rangle)^T\right)^T$. Appling a proper dephasing map \cite{L-L} 
to this output as ${\cal D}(\rho'_{out})$, we will obtain a direct sum of $A_i\rho_{in}A_i^{\dagger}$ in ${\cal H}_T$, which can be transformed by a contraction map ${\cal L}$ 
\cite{reason3} to ${\cal W}(\rho_{in})=(\sum_{i=1}^n A_i\rho_{in}A_i^{\dagger})/n$ in one subspace of ${\cal H}_T$. 
The output state of the total map ${\cal W}$ on a pure state input $|\psi_{in}\rangle$, normalized as in Eq. (3), is 
$\rho_{out}=\sum_i A_i\rho_{in}A_i^{\dagger}$ of a QO mapping $|\psi_{in}\rangle$ from a pure to mixed state.
A general QO can be therefore realized by a corresponding combined map ${\cal W}$. 
To a single set of detectors that effect a dephasing map by detecting $A_i|\psi_{in}\rangle$ coming from different terminals, 
what is being measured is effectively the output of a general QO (see the caption of Fig. 3).

In conclusion, we have presented the linear optics schemes (including the photon detection) to realize all QOs and POVMs on a 
single photon {\it qudit}. The circuits to perform all the relevant tasks are only the combinations of some scalable unitary operator modules which have been 
widely applied in quantum information processing. Given current technologies, our schemes can realize any permissible map on single photon signals in a deterministic way. 

\begin{acknowledgments}
B.H. and J.B acknowledge the partial support by a grant of PSC-CUNY.
Z.W. thanks the support of the National Natural Science Foundation of China (Grant No. 60671030).
\end{acknowledgments}

\bibliographystyle{unsrt}

\begin{thebibliography}{99}
\bibitem{p-m} P. Kok, W. J. Munro, K. Nemoto, T. C. Ralph, J. P. Dowling and G. J. Milburn, Rev. Mod. Phys. {\bf 79}, 135 (2007).
\bibitem{cryto}N.\ Gisin, G.\ Ribordy, W.\ Tittel and H. Zbinden, Rev. Mod. Phys. {\bf 74}, 145 (2002).
\bibitem{reck}M.\ Reck, A.\ Zeilinger, H.\ J.\ Bernstein and P.\ Bertani, Phys. Rev. Lett. {\bf 73}, 58 (1994).
\bibitem{torma96} P.~T{\" o}rm{\" a} and S.~Stenholm,
\newblock Phys. Rev. A {\bf 54}, 4701 (1996).
\bibitem{torma95} P.~T{\" o}rm{\" a}, S.~Stenholm and I.~Jex,
\newblock Phys. Rev. A {\bf 52}, 4853 (1995).
\bibitem{jex95} I.~Jex, S.~Stenholm and A.~Zeilinger, \newblock Opt. Commun. {\bf 117}, 95 (1995).
\bibitem{AP93} See, e.g., A.~Peres, {\it Quantum Theory: Concepts and Methods} (Kluwer
Academic Publishers, Dordrecht, The Netherlands, 1993).
\bibitem{cal}See, e.g., J.\ Calsamiglia, Phys. Rev. A {\bf 65}, 030301(R) (2002).
\bibitem{bt}See, e.g., R.  Bhatia, {\em Matrix Analysis}, (Springer, New York, 1997).
\bibitem{bibkraus}K. Kraus, {\em Lecture Notes: States, Effects and Operations}, (Springer, New York, 1983).
\bibitem{input}The input does not couple to the environment $\rho_{E}$ or any other system to evolve as the tensor product 
$\rho_{in}\otimes\rho_{E}$. For photon states, their coupling with the environment is negligibly weak.
\bibitem{d-p} G. M. D'Ariano and P. Lo Presti, \prl {\bf 86}, 4195 (2001).
\bibitem{b-d-s}F.\ Buscemi, G.\ M.\ D'Ariano and M.\ F.\ Sacchi, Phys. Rev. A {\bf 68}, 042113 (2003).
\bibitem{definition}For the Hermitian operators $\Pi_i$, we simply have 
$||\Pi_i||=max\{|\lambda_k|: \Pi_i|\phi_k\rangle=\lambda_k|\phi_k\rangle, k=1,\cdots, N \}$.
\bibitem{reason1}Since $||UAV||=||A||$ for an arbitary linear operator $A$ and all unitary matrices $U$ and $V$, we obtain the following from Eq. (12):
$$||\Sigma_2^{\ast}||=||\Sigma_2^{\ast }U_{2L}U_1||
 = ||\Pi_2^{1/2}(I-\Pi_1)^{-1/2}||\leq 1, $$
where we have used Lemma V.1.7 in \cite{bt} with the existence of $(I-\Pi_1)^{-1}$ due to the fact that $||\Pi_1||<1$.
\bibitem{reason2}Suppose one of the POVM elements, $\Pi_j$, has a unit norm, so its diagonalized form $\Sigma^2_j$ has some entries
$1$. From Eq. (1), on the other hand, we have
$$\Sigma^2_j+\sum_{i\neq j}U_jU^{\dagger}_i\Sigma^2_{i}U_iU^{\dagger}_j=I,$$
and then we find that the corresponding entries of $\Sigma^2_i$, for all $i\neq j$, are $0$.
\bibitem{ap}S.\ E.\ Ahnert and M.\ C.\ Payne, Phys. Rev. A {\bf 71}, 012330 (2005).
\bibitem{L-L}P.\ van Loock and N.\ L{\" u}tkenhaus, Phys. Rev. A {\bf 69}, 012302 (2004).
\bibitem{reason3} If $n=2$, for example, this linear contraction map can be chosen, e.g., as the following ($O$ represents a block with 0 entries):
$${\cal L}=\frac{1}{\sqrt{2}}\left(\begin{array}{cc}O& O\\
 I& I\\
\end{array}\right),$$
with $||{\cal L}||=1$. The combined map, 
$${\cal W}(\rho_{in})={\cal P}{\cal L}{\cal D}\bigl({\cal U}\rho_{in}{\cal U}^{\dagger}\bigr){\cal L}^{\dagger}{\cal P}=\frac{1}{2}(A_1\rho_{in}A_1^{\dagger}+A_2\rho_{in}A_2^{\dagger}),$$
with ${\cal P}$ being the projection onto a subspace, outputs a mixed state.
\bibitem{watrous} See, e.g., J.\ Watrous, CPSC 701 Lecture Notes: {\em Theory of Quantum Information}, University of Waterloo.
\end{thebibliography}

\end{document}